# Rydberg atoms in hollow-core photonic crystal fibres

G. Epple[1,3], K. S. Kleinbach[3], T. G. Euser[1], N. Y. Joly[1,2], T. Pfau[3], P. St.J. Russell[1,2] and R. Löw[3]

[1]Max Planck Institute for the Science of Light and [2]Department of Physics, University of Erlangen,

Günther-Scharowsky-Str. 1, 91058 Erlangen, Germany

[3]5. Physikalisches Institut, Universität Stuttgart, Pfaffenwaldring 57, 70550 Stuttgart, Germany

**The exceptionally large polarisability of highly excited Rydberg atoms – six orders of magnitude higher than ground-state atoms – makes them of great interest in fields such as quantum optics[1-3], quantum computing[4-6], quantum simulation[7] and metrology[8]. If however they are to be used routinely in applications, a major requirement is their integration into technically feasible, miniaturised devices. Here we show that a Rydberg medium based on room temperature cæsium vapour can be confined in broadband-guiding kagomé-style hollow-core photonic crystal fibres. Three-photon spectroscopy performed on a cæsium-filled fibre detects Rydberg states up to a principal quantum number of $n = 40$. Besides small energy level shifts we observe narrow lines confirming the coherence of the Rydberg excitation. Using different Rydberg states and core diameters we study the influence of confinement within the fibre core after different exposure times. Understanding these effects is essential for the successful future development of novel applications based on integrated room temperature Rydberg systems.**

Hollow-core photonic crystal fibre (HC-PCF) has opened up new opportunities in the field of atomic spectroscopy, enabling both light and atoms to be confined in the same narrow channel over distances that exceed the Rayleigh length by many orders of magnitude.[9] Filled with alkali atoms, HC-PCF has been used to study low-lying excitations in ultracold alkali gases[10] as well as in thermal vapours.[11-13] Nonlinear optical experiments in particular benefit from the enhanced overlap between light and matter provided by HC-PCF, a recent example being the demonstration of all-optical switching by electromagnetically-induced transparency using only a few hundred gating photons.[10-12]

The large mutual interactions between Rydberg states can provide even stronger optical nonlinearities.[1] Their origin lies in the extreme polarisability of Rydberg states. This leads to strong dipole-dipole or van-der-Waals interactions, which are effective over micron-scale distances even at room temperature.[14] One consequence of this is the Rydberg excitation blockade, which is the key element for controlled Not gates[4,6], single photon sources[2], antibunched[3] and attractive photons[15], single photon absorbers[16] and atom-light entanglement[17].

Every quantum optics experiment involving Rydberg states requires excellent control over the dimensions and arrangement of the atoms. Of particular interest is a one-dimensional configuration, where the blockade radius (typically a few microns) is larger than the width of the excitation volume. Such systems have recently been realised in free-space cold atom experiments, where the generation of anti-bunched[3] and attractive photons[15] has been successfully demonstrated. HC-PCF is a particularly suitable system[18], especially for thermal atoms, as it provides optically accessible, easy-to-fill, micron-scale confinement volumes capable of withstanding corrosive alkali vapours. Additionally, the effective length of the one-dimensional system, which in free space experiments is limited by diffraction, can be significantly increased. For successful integration of room temperature Rydberg systems into micron-scale devices, however, it is essential to understand how the Rydberg atoms are influenced by the nearby glass walls of the core.[19]

While some effects, such as local fields created by adsorbed ions, mainly depend on the distance between the atom and the glass-wall, other interactions are also affected by the thickness of the glass-wall itself, e.g., Casimir-Polder effects. These interactions are expected to be weaker in HC-PCF (compared to glass cells or capillaries) owing to the greatly reduced thickness of the glass walls surrounding the core (typically ~200 nm thick).

Here we want to demonstrate the feasibility of exciting Rydberg atoms inside a HC-PCF as well as first results of effects due to the fibre confinement. Therefore we have studied the influence of different core diameters as well as different optical densities on to a wide range of Rydberg states, so as to identify an optimal design for future integrated Rydberg systems based on HC-PCF.

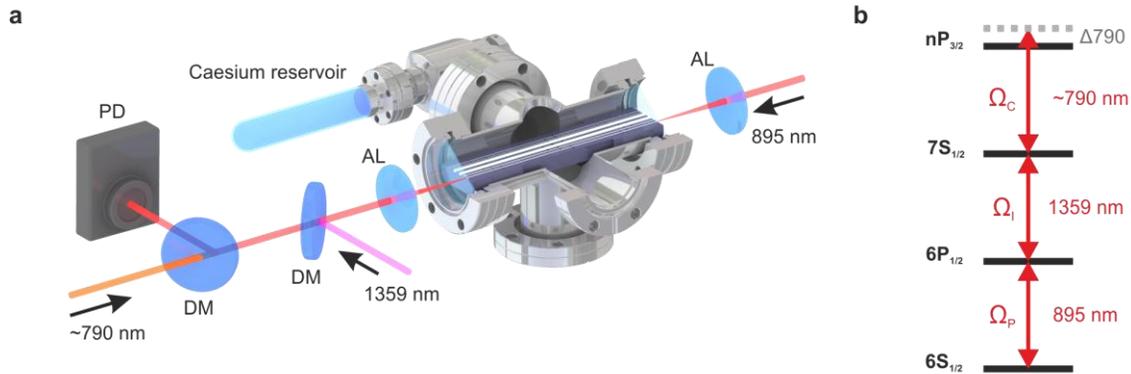

Figure 1: **Experimental set-up and excitation scheme.** (a) A selection of kagomé PCFs with different core diameters is mounted inside a vacuum chamber containing saturated cæsium vapour. The three excitation beams are coupled into the fibre from both sides using achromatic lenses (AL). The beams are superimposed/separated using dichroic mirrors (DM), and the transmission of the counter-propagating 895 nm probe beam is measured at a photodiode (PD). (b) Three-photon excitation scheme up to a Rydberg level. The lasers driving the two lower transitions are locked on resonance while the Rydberg transition laser at 790 nm is scanned in frequency.

In the experiments we used two different kagomé-style HC-PCFs (kagomé-PCFs), cleaved to lengths of ~13.5 cm and mounted in a UHV chamber, as shown in the experimental set-up in Fig. 1**a**. Scanning electron micrographs of the cross-sectional structure of both fibres are shown in Fig. 2; the core diameters are ~60 μm and ~19 μm. For the Rydberg spectroscopic measurements a coherent three-photon excitation scheme[20] involving lasers at 895, 1359 and 790 nm was chosen (Fig. 1**b**). The fibres were specifically selected to guide these wavelengths. All the beams were coupled into the fundamental mode of the fibre by focusing the light through UHV-windows onto the fibre ends.

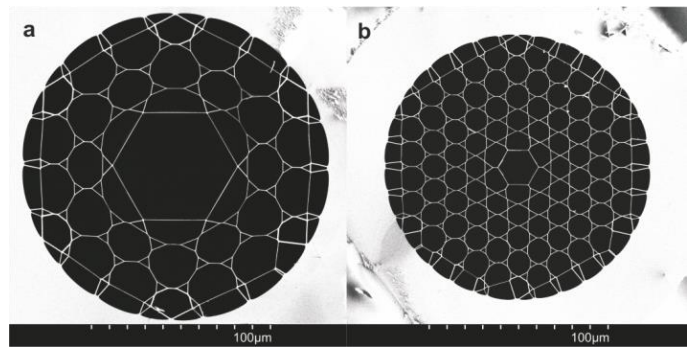

Figure 2: **Scanning electron micrographs (SEMs) of the two kagomé-PCFs studied.** The special kagome-lattice cladding provides broadband transmission. The diameter of the hollow core region is ~60 μm for fibre (a) and ~19 μm for fibre (b).

The transmitted 895 nm signal served as a probe and its power was recorded by a photodiode. To compensate partly for the Doppler-effect, the other two beams were launched into the fibre in the opposite direction. The polarisation of all three beams was linear and parallel. The frequencies of the 895 and 1359 nm lasers were locked to sub-Doppler accuracy using a reference cell, while the laser driving the coupling transition at 790 nm was scanned in frequency. For comparison we always simultaneously recorded a reference spectrum in a conventional 5 mm thick glass cell. The spectroscopy scheme was identical to that in the fibre.

As most of the properties of Rydberg atoms scale, in varying ways, with the principal quantum number $n$, we performed measurements at different values of $n$. A selection of the spectra obtained in the 60 μm core fibre is plotted in Fig. 3**b** and **c** along with the reference signal (Fig. 3**a**). The principle quantum number $n$ of the excited Rydberg states (all of them of type $nP_{3/2}$) was increased from 28 via 34 to 40. To correct for the influence of the inner electrons, which are absent in a hydrogen atom, $n$ is replaced by the effective principal quantum number $n^* = n - \delta_P$ where $\delta_P = 3.558$ is the quantum defect[21]. The origin of the frequency axis is determined by the reference signal position. To avoid any change in atomic density within one measurement series due to light-induced desorption[22] or heating of the fibre, the laser power was kept constant. As a consequence the observed coupling of the light fields to the Rydberg states decreased with increasing $n^*$, resulting in a weaker signal.

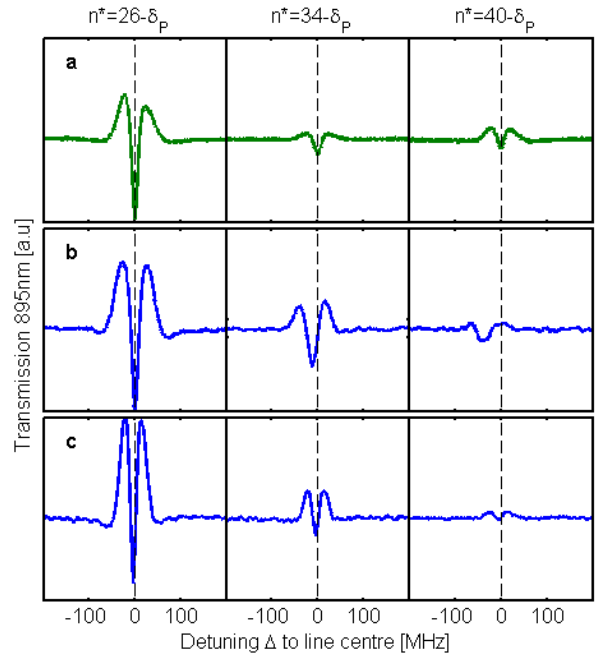

Figure 3: **Coherent three-photon spectroscopy for different Rydberg states.** Transmission spectra of the probe light for the reference cell (a), defining the origin of the frequency axes, and for the 60 μm core with an OD=0.4 (b) and OD=2.9 (c) relating to exposure times of two (b) and four (c) month. The signal is shown for three different

principle quantum numbers $n^* = n - \delta_P$. The decreasing signal strength for higher values of $n$ originates from a decrease in the magnitude of the dipole matrix elements. For the lower OD (b) and higher $n$ a frequency shift of the signal in fibre relative to the reference cell signal is observable, which we attribute to background electric fields. For the higher OD (c) this shift is not observable anymore. All measurement parameters are given in the Methods section.

The measurements inside the fibre were carried out at different optical densities, in this case defined as OD=$-\ln(I_{out}/I_{in})$ at the D1 line (F=3 to F'=4) resonance. The optical density was calculated to be OD=0.4 for the measurement series in Fig. 3**b** and OD=2.9 for Fig. 3**c** respectively (see Methods). Since both experiments were carried out at room temperature we mainly attribute this difference to the longer exposure time of the fibre to the background cæsium vapour of four months (Fig. 3**c**) compared to two months (Fig. 3**b**). But also heating of the chamber as well as shining in intense laser fields during the time between the experiments might have had additional effects on the atomic density inside the fibre. The strongest deviation between the reference and the fibre signal is observable for high $n^*$ at the smaller OD (Fig. 3**b** right-hand side), were the signal position is clearly shifted to lower frequencies, indicating the Rydberg states are influenced by the confining environment.

However for the measurements at higher OD this effect is not observable any more (Fig.3**c**). In this case the OD in fibre and the surrounding chamber (OD=2.4) are nearly the same. Therefore we conclude that we have finally reached an equilibrium situation with homogeneous filling of the fibre. Nevertheless, the question remains by what the line shifts were caused in the first place and why they disappeared finally.

To obtain a better understanding of possible influences onto the Rydberg atoms we systematically measured this signal shift for a larger set of Rydberg states. In Fig. 4 the frequency shift between the signal inside the 60 μm fibre (blue squares) and the reference signal position is plotted versus the effective quantum number $n^*$ for both optical densities OD=0.4 (filled blue squares) and OD=2.9 (open blue squares). For the lower OD a clear shift is observable which increases as we go to higher $n^*$. To identify the origin of this shift $\delta$ it is useful to compare its behaviour with well-known scaling laws for Rydberg atoms $(n^*)^m$.[23] Fitting the experimental data to a power law as shown in the log-log scaled inset in Fig. 4 results in $m = 6.7\pm0.7$ (solid blue line). Since the polarisability $\alpha$ of Rydberg atoms scales with $m = 7$, we are confident that the shift in the 60 μm is predominantly caused by static electric fields. By interpolating known values of polarisabilities for cæsium $P_{3/2}$ Rydberg states[24] we estimate the field strength, using $\delta = \alpha E^2/2$, to be ≈0.5 V/cm (Methods). In contrast to that the shift was not observable anymore for the higher OD even at high $n$.

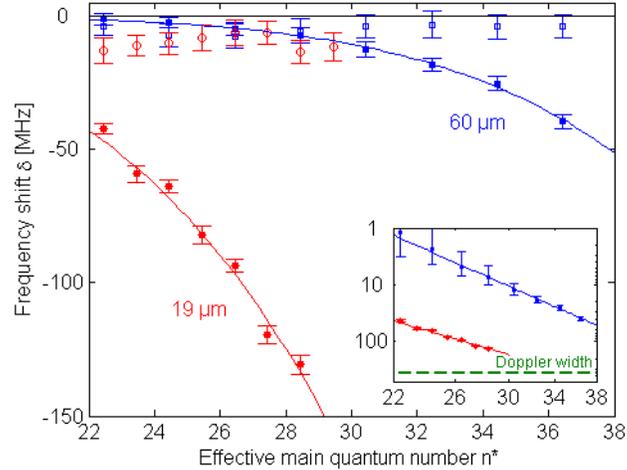

Figure 4: **Signal shift inside fibre.** The frequency shifts between fibre and reference cell signal are shown as a function of the effective main quantum number for the 60 μm fibre (blue squares) and the 19 μm core fibre (red dots). The measurements were performed at exposure times of two month (filled) and around four month (open) resulting in ODs of 0.4/2.9 in the 60 μm fibre and 0.2 /1.5 in the 19 μm fibre. Fitting the shift observable after two month of exposure time to $\delta \sim (n^*)^m$ power laws leads to $m = 6.7\pm0.7$ for the 60 μm core fibre and $m = 4.4\pm0.4$ for the 19 μm core fibre plotted on a log-log scale in the inset where the Doppler line width is indicated by the horizontal dashed line for comparison.

Additionally we investigated the influence of a reduced mean distance between the Rydberg atoms and glass walls by measuring in a 19 μm core fibre (red dots). The exposure times are comparable to the ones in the 60 μm core fibre, in this case resulting in optical densities of OD=0.2 (filled red dots) and OD=1.5 (open red dots). For the lower OD we again observed an even larger redshift of the signal. The fitted power law shown in the inset leads to $m = 4.4 \pm 0.4$ (solid red line). This shift in the 19 μm core does not match the scaling behaviour for electric fields and is not fully understood yet. It is important to point out that for all measurements the frequency shift of the signal is smaller than the FWHM Doppler line width of ≈360 MHz represented by the horizontal dashed line in the inset in Fig. 4.

The OD=1.5 measured after around four months of exposure time inside the 19 μm core fibre (compared to an OD=2.4 outside) indicates that due to the smaller core size and therefore longer diffusion time, equilibrium has not been reached yet. Nevertheless, also in the 19 μm fibre for this higher OD no systematic shift was observable even for high $n^*$ (open red dots).

This may be explained by a decreased amount of nearby charges or due to higher and more homogeneous cæsium density in fibre, leading to an improved shielding of any electric fields. However, in both fibres the shifts have significantly been reduced giving an even more favourable situation for any applications.

The results demonstrate that the spectroscopy of Rydberg states inside HC-PCF is not fundamentally limited by interaction with the confinement at the moment. However, when approaching the goal of a one dimensional system, which requires even smaller core sizes or if higher excited Rydberg states are requested we expect again line shifts due to resonant or off-resonant coupling to the material like e.g. Casimir polder forces. Further systematic measurements in combination with an improved spectroscopy resolution will be needed to investigate these regimes and gain a more complete understanding of confined Rydberg atoms. Nevertheless the results presented here are already now highly promising, opening up various possibilities. For example, the selective excitation of higher-order modes[25] in fibre will allow atoms to be probed at different distances from the core walls. Higher order modes, such as the $LG_{10}$, can also provide even tighter radial confinement of the light field, permitting the diameter of a blockaded system to be further reduced towards a genuinely one-dimensional system.

But also Rydberg gases in the non-interacting regime enable many interesting applications. For example, fibre-based spectroscopy cells could be used as non-invasive microscopic microwave detectors.[8] Additional functionality could be achieved by incorporating gold nanowires[26] into the fibres, either to enable direct detection of the Rydberg population via a current[27], or to frequency-modulate the probe light[28] at extremely low electric fields. Ultimately, splicing atom-filled HC-PCF to standard single mode fibres[29] could allow integration of systems based on Rydberg atoms into fibre-coupled devices operating at room temperature.

## Methods

The optical depth inside the fibre was calculated by an absorption measurement of the 895 nm light scanning the D1 line. To ensure that the spectroscopy signal was dominated by atoms inside the hollow-core the absorption signal with 895 nm light coupled into the fundamental core mode was compared to the situation where the 895 nm light was coupled into the glass cladding. Since the ODs measured when light is coupled into the cladding are at least one order of magnitude smaller this assumption is valid. To extract the final in fibre ODs the cladding values were subtracted from the core values.

The laser systems for 895 nm and 1359 nm light were frequency locked to the $6S_{1/2} \rightarrow 6P_{1/2}$ transition (D1 line) by Doppler-free spectroscopy and to the $6P_{1/2} \rightarrow 7S_{1/2}$ transition by polarisation spectroscopy. The three beams were superimposed using dichroic mirrors and launched into the kagomé-PCF with achromatic lenses.

To achieve a better signal to noise ratio a lock-in amplifier was used. Therefore, in the first measurement series (two months exposure time) the 1359 nm light was modulated at 1.1 kHz by a chopping wheel. Further measurements showed that if the 790 nm light is modulated fast enough, no influence of the OD due to LIAD or heating effects is observable for powers used in the experiment. Therefore in the second measurement series (four months exposure time) the setup was updated with now modulating the 790 nm light at 50 kHz via an AOM.

The launch efficiency into the fibre was optimised both by imaging the mode profile at the end of the fibre and optimising the transmitted power. This way we ensured that most of the light was coupled into the fundamental fibre mode for all beams. Experimental launch efficiencies of ~30% (1359 nm), ~40% (895 nm) and ~60% (790 nm) were achieved for the 60 μm core, and ~20% (1359 nm), ~30% (895 nm) and ~40% (790 nm) for the 19 μm core. These values include transmission losses in the fibre and the vacuum window. The input powers in the first measurement series (after around two months exposure time) for the 60 μm (19 μm) fibre were $P_P = 0.15$μW (0.3 μW) for the 895 nm probe beam, $P_I = 7.0$ μW (6.0 μW) for the 1359 nm intermediate beam and $P_C = 70$ mW (50 mW) for the 790 nm coupling beam. Analogous the values for the measurement series after around four months were $P_P = 0.5$μW (0.15 μW), $P_I = 10.0$ μW (10.0 μW) and $P_C = 30$ mW (20 mW) for the 60 μm (19 μm) fibre. For the measurements in the reference cell shown in Fig. 3 the adjusted powers were $P_P = 2$ μW, $P_I = 200$ μW and $P_C = 20$ mW, and the beam radii were $w_P = 41.1 \pm 1.0$ μm, $w_I = 64.3 \pm 1.2$ μm and $w_C = 46.3 \pm 1.6$ μm. The signals are averaged between 100 and 300 times.

The $1/e^2$ beam radius in the fibre core were calculated to be ~20 μm for the 60 μm core fibre and ~12 μm for the 19 μm core fibre. The transit time broadening for the different systems can be calculated to be $2\pi \times 0.7$ MHz for the reference cell, $2\pi \times 1.5$ MHz for the 60 μm core fibre and $2\pi \times 4.9$ MHz for the 19 μm core fibre. Since the signal itself is broader we were not limited by transit-time effects. This broadening mechanism will however become more important when the core size is further reduced.

The raw data was evaluated in the following way: The relative frequency change of the coupling laser (*x*-axis in Fig. 3) was measured using a Fabry-Pérot interferometer. The detuning zero was extracted by fitting a quadratic function to the absorption dip at line-centre in the reference signal. The central frequency of the fibre signal was located in the same manner. Since only the fibre signal was fed through a lock-in amplifier, one need to consider the time delay caused by the electronics, that will translate into a constant offset frequency shift between fibre and reference signal. This delay was measured to be ~ 8.0 ms for the first measurement series and ~ 400 μs for the second depending on the different integration times. In future experiments both signals will be fed through a lock in amplifier of the same typ. The error bars on the frequency shift in Fig. 4 include the uncertainty in the lock-in amplifier delay, the uncertainty in scaling the frequency axis and uncertainties in fitting the line-centres.

Acknowledgment:

This project has been financed by the Baden-Württemberg Stiftung and supported by the ERC under contract number 267100. G.E. acknowledges support from the International Max Planck Research School "Physics of Light"